\documentstyle[prb,aps,epsf,floats,twoside]{revtex}  
\begin{document}

\twocolumn[\hsize\textwidth\columnwidth\hsize\csname
@twocolumnfalse\endcsname

\draft

\title{Magnetic Field-Induced Fano Profiles in the Absorption
Coefficient of Semiconductors}

\author{V. I. Belitsky, A. Cantarero, and S.T.
Pavlov\cite{Pav}}
\address{Departamento de F\'{\i}sica Aplicada, Universidad de
Valencia, Burjasot, E-46100 Valencia, Spain}

\date{\today}

\maketitle

\begin{abstract}
A strongly asymmetric, Fano-like resonance profile has been found for
magneto-absorption in the states of hot 
free electron-hole pairs scattered by
defects in bulk semiconductors. The renormalization 
of the absorption profile, compared to that expected from the bare density
of states in a high magnetic field,  
follows from the quasi-one-dimensional character
of electronic excitations. The results are valid for absorption
by the electronic states of large Landau quantum number 
where the Coulomb interaction should 
play a minor role. The resonance shape is in a good 
qualitative agreement with experimental observations.   
\end{abstract}
\vskip 2pc ]
\narrowtext

Magneto-absorption spectra\cite{hl,ul,g1}                                         
of some bulk semiconductors  
displays strongly asymmetric Fano-like resonances.       
In a recent paper of Glutsch {\em et al.}\cite{g1} the observed 
profiles have been explained by
coupling between the discrete and continuum states of
magneto-excitons. The authors show that the necessary coupling
may be induced by Coulomb interaction. Both experimental
and theoretical profiles 
obtained have asymmetric form. However, the higher energy dip of the
resonance seems to be much more pronounced in the 
data than in the results of the calculations.   Moreover, the role of 
Coulomb interaction between electron  
and hole should be strongly weakened 
in the high Landau number $N$ range where the magnetic energy 
$N\hbar\omega_c$  
is large compared to the excitonic Rydberg. 

We suggest here that the
qualitative explanation of the Fano profile in 
magneto-absorption stems from the quasi-one-dimensional
character of the electronic excitations, 
and the corresponding singularity in
the density of states (placed into continuum of lower Landau
bands), rather than from the details of
the interaction. We show that a very strong asymmetry,
qualitatively resembling 
Fano profiles, can be found in the magneto-absorption by
uncorrelated electron-hole pairs when {\em strong enough elastic 
scattering} by 
impurities or other defects is assumed be the relevant 
mechanism. An exact analytical solution of the problem seems to
be impossible, and even very advanced numerical calculations face 
serious problems as it was demonstrated in
Ref.~[\onlinecite{g1}~] for absorption by magneto-excitons. 
Our approach is not aimed at a 
quantitative analysis of the data but rather at remarking {\em  
qualitatively} the possible effect of scattering 
on the absorption by quasi-one-dimensional
electronic excitations in a high magnetic field. 
Since free electron-hole pairs are 
considered below, the results may break down in the energy
interval close to the fundamental gap. 

Neglecting the electron-hole interaction, 
the absorption coefficient $\alpha$ can be written in a
factorized form      
\begin{eqnarray}
\label{1}
\alpha &=&{4n(\omega_l)e^2|{\bf e}_l{\bf p}_{cv}|^2\over
V_0m_0^2c\hbar\omega_l} \int d\omega\sum_{N,k_y,k_z}\nonumber\\
&&{\rm Im}G_e(N,k_z;\omega)
{\rm Im}G_h(N,-k_z;\omega_l-\omega)~,
\end{eqnarray}
where $\omega_l$ (${\bf e}_l$) 
is the incident light frequency (polarization), 
$n$ the refractive index, $e$ the electron charge, ${\bf
p}_{cv}$ is the interband matrix element of the momentum
operator, $V_0$ is the normalization volume, and $m_0$ the bare electron
mass.  
$G_e$ ($G_h$) is the electron (hole) Green function which
is renormalized, in principle, by all relevant interactions;    
when taken as a solution of the Dyson equation, it gives 
\begin{eqnarray}
\label{2}
&&~~~~\alpha={4n(\omega_l)e^2|{\bf e}_l{\bf p}_{cv}|^2\over
V_0m_0^2c\hbar\omega_l}\int d\omega\sum_{N,k_y,k_z}\nonumber\\
&&{\rm
Im}\left[G_{0e}^{-1}(N,k_z;\omega)-\Xi_e(N,k_z;\omega)
\right]^{-1}\\
&\times&~{\rm Im}\left[G_{0h}^{-1}(N,-k_z;\omega_l-\omega)
-\Xi_h(N,-k_z;\omega_l-\omega)\right]^{-1},\nonumber
\end{eqnarray}
where $G_0$ is a bare Green function and $\Xi$ the self-energy
evaluated below in the lowest order of perturbation theory.  

To show how the self-consistent treatment of
scattering in the lowest order perturbation theory 
results in a dramatic change of the absorption
profile, we use a simple model of a parabolic band for electrons 
and {\em infinitely heavy holes}. The latter assumption will
enable us to integrate analytically Eq.~(\ref{2}).   
The scattering of the electrons is 
modeled by a $\delta$-function interaction with point defects. 
Evaluation of the 
lowest order contribution to the self-energy gives immediately    
\begin{equation}
\label{3}
\Xi_e(\omega)=-i\Lambda\sum_{N^{\prime}}{(\hbar\omega_{ce})^{3/2}
\over\sqrt{\hbar\omega  
-(N^{\prime}+1/2)\hbar\omega_{ce}}}~,
\end{equation}
where  
$$\Lambda =v^2n_{imp}\hbar\omega_{ce}{m_e^{3/2}
\over\pi 2^{3/2}\hbar^3}$$ 
is a dimensionless coupling proportional 
to the square of the scattering potential strength 
$v$ and to the impurity
concentration $n_{imp}$, $m_e$ is the electron effective mass.
Note that the idealized $\delta$-function form of the scattering
potential by defects results in the divergence of the real part
of self-energy. To calculate the profiles we cut the number of
terms in the sum over Landau index to a reasonable value
determined by the laser frequency and the magnetic field
strength. 

\begin{figure}[t]
\vspace{-3cm}
\epsfxsize=3 in
\epsffile{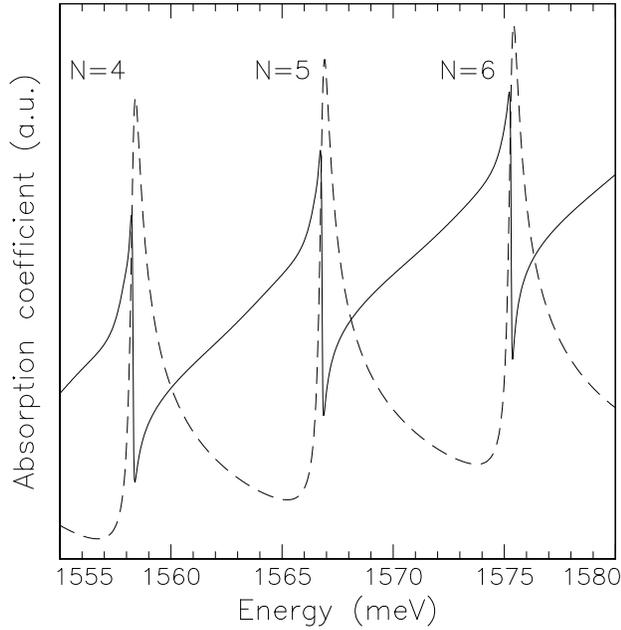}
\vspace{1cm}
\caption{The results of calculations for magneto-absorption
obtained for $H=5.2~T$ and according to the model discussed in
the text. The gap is 1.52 eV, effective mass of the electron 
$m_e=0.067~m_0$, and $\Lambda=0.2$. The dashed line shows
the profile calculated for bare Landau levels with a constant
broadening of $0.02~\hbar\omega_{ce}$.
\label{2f}}
\end{figure}

Under the assumptions made,   
the absorption coefficient reduces to 
\begin{eqnarray}
\label{7}
&&~~~\alpha=-{2m_en(\omega_l)e^2|{\bf e}_l{\bf p}_{cv}|^2\omega_{ce}\over
\pi m_0^2c\hbar\omega_l}\sum_N\int dk_z\nonumber\\
&&{\rm
Im}\left[\hbar\omega_l-\hbar\omega_g-{\hbar^2k_z^2\over 2m_e}-\left(N
+{1\over 2}\right)\hbar\omega_{ce}\right.\nonumber\\
&+&i\Lambda\left.
\sum_{N^{\prime}}{(\hbar\omega_{ce})^{3/2}\over\sqrt{
\hbar\omega_l-\hbar\omega_g-(N^{\prime}+1/2)\hbar
\omega_{ce}}}\right]^{-1}~.
\end{eqnarray}
Equation (\ref{7}) can be easily integrated 
analytically to yield: 
\begin{eqnarray}
\label{4}
&&\alpha ={2\sqrt{2}n(\omega_l)e^2m_e^{3/2}
|{\bf e}_l{\bf p}_{cv}|^2\omega_{ce}\over
m_0^2c\hbar^2\omega_l}\nonumber\\
&&\times{\rm Re}\sum_N\left[
\hbar\omega_l-\hbar\omega_g-\left(N+{1\over 2}\right)
\hbar\omega_{ce}\right.\nonumber\\
&&\left.+i\Lambda\sum_{N^{\prime}}{(\hbar\omega_{ce})^{3/2}
\over\sqrt{\hbar\omega_l
-\hbar\omega_g-(N^{\prime}+1/2)\hbar\omega_{ce}}}\right]^{-1/2}. 
\end{eqnarray}
The results of calculation according to
Eq.~(\ref{4}) and for coupling $\Lambda=0.2$ 
are shown in Fig.~\ref{2f}. We want to emphasize that, due
to the resonant character of the renormalization   
(see Eq.~(\ref{3})), the minimum point on the higher energy side of
each resonance {\em corresponds to the singularity in the bare
density of states for  Landau level $N$}. 
However, every individual resonance in
Eq.~(\ref{4}) is 
composed of contributions from different Landau
levels lying  
{\em both above and below} the bare level $N$. The resulting shape of the
resonance is very different from the bare density of
states and magneto-absorption profile acquires the Fano-type
form.

Our model involves essentially only the electron density of
states since the infinite effective mass of the hole have 
been assumed in calculations. A more elaborate theory is needed for 
quantitative analysis of the data.

\acknowledgements
The authors thank Prof. M. Cardona and Dr. T. Ruf for
indicating the problem and 
stimulating discussions, and Prof. R. G. Ulbrich for useful comments. 
V. I. B. and S. T. P. thank the European Union,  
Ministerio de Educacion y 
Ciencia de Espa\~na (DGICYT) and the Russian Fundamental 
Investigation Fund (93-02-2362) 
for financial support and the 
University of Valencia for its hospitality. This work has been 
partially supported by Grant PB93-0687 (DGICYT).

\end{document}